\documentclass[a4paper,11pt,dvips]{article}
\usepackage{graphicx}
\usepackage{amsmath}
\usepackage{amssymb}
\usepackage{latexsym}
\usepackage[colorlinks]{hyperref}
\usepackage{color}
\usepackage{float}
\usepackage{cite}
\usepackage{makeidx}
\usepackage{colortbl}

\newcommand{\bra}{\begin{array}}
\newcommand{\era}{\end{array}}
\newcommand{\beq}{\begin{equation}}
\newcommand{\eeq}{\end{equation}}
\newcommand{\beqar}{\begin{eqnarray}}
\newcommand{\eeqar}{\end{eqnarray}}

\def\BC{\bb C}
\def\_\BC{\bbi C}

\def\( {\left(}
   \def\) {\right)}
\def\[ {\left[}
\def\] {\right]}
\def\no2 {{\textstyle{n\over 2}}}

\def\dag {{\dagger}}

\newcommand{\lb}{\label}

\begin{document}
\begin{titlepage}
\setcounter{page}{1}
\renewcommand{\thefootnote}{\fnsymbol{footnote}}

\begin{flushright}
ucd-tpg:1103.05\\
\end{flushright}

\vspace{5mm}
\begin{center}

{\Large \bf {Transmission through  Biased Graphene Strip}}

\vspace{5mm}

{\bf H. Bahlouli}$^{a,b}$, {\bf E.B. Choubabi}$^{a,c}$,
{\bf A. El Mouhafid}$^{a,c}$ and
{\bf A. Jellal\footnote{\sf ajellal@ictp.it -- jellal.a@ucd.ac.ma}}$^{a,c,d}$

\vspace{5mm}

{$^a$\em Saudi Center for Theoretical Physics, Dhahran, Saudi Arabia}

{$^b$\em Physics Department,  King Fahd University
of Petroleum $\&$ Minerals,\\
Dhahran 31261, Saudi Arabia}

{$^{c}$\em Theoretical Physics Group,  
Faculty of Sciences, Choua\"ib Doukkali University},\\
{\em PO Box 20, 24000 El Jadida,
Morocco}

{$^d$\em Physics Department, College of Sciences, King Faisal University,\\
PO Box 9149, Alahsa 31982, Saudi Arabia}


\vspace{3cm}

\begin{abstract}

We solve the 2D Dirac equation describing graphene in the presence of a linear vector potential.
The discretization of the transverse momentum due to the infinite mass boundary condition reduced our
2D Dirac equation to an effective massive 1D Dirac equation with an effective mass equal to the quantized
transverse momentum. We use both a numerical Poincar\'e Map approach, based on space discretization of the
original Dirac equation, and direct analytical method. These two approaches have been used to study tunneling
phenomena through a biased graphene strip. The numerical results generated by the Poincar\'e Map are
in complete agreement with the analytical results.

\vspace{3cm}



\end{abstract}
\end{center}
\end{titlepage}

\section{Introduction}

Graphene, a single layer of carbon atoms laid out in a honeycomb lattice, is one of the most
interesting electronic systems discovered in recent  years \cite{ag,ah}. It differs from conventional
two dimensional electron gas (2DEG) systems in that the low energy physics is governed by a massless
Dirac Hamiltonian rather than the more common form used for semiconductors, characterized by an effective
mass and a band gap.

The tunneling of Dirac fermions in graphene has  already been verified experimentally~\cite{n},
 which in turn has spurred an extraordinary amount of interest in the investigation of the electronic
 transport properties in graphene based quantum wells, barriers, $p$-$n$ junctions, transistors,
 quantum dots, superlattices,
$\cdots$ etc. The electrostatic barriers in graphene can be generated in various ways~\cite{ka,hs},
by applying a gate voltage, cutting it into finite width nanoribbons and using doping or otherwise.
On the other hand, magnetic barrier could in principle 
be realized with the creation
of magnetic dots. In the case of graphene, results of the transmission coefficient and the tunneling
conductance were already reported for the electrostatic barriers{~\cite{ka,hs,ma,s,ben,ad,nov}}
and
{magnetic barriers~\cite{mr,ch,mo}}.

The fact that in an ideal graphene sheet the carriers are massless gives rise to Klein paradox,
 which allows particles to tunnel through any electrostatic potential barriers, that is the wavefunction
 has an oscillatory tail outside the electrostatic barrier region. Hence this property excludes the possibility
 to confine electrons using electrostatic gates, as in usual semiconductors. Thus to enable the fabrication of
 confined structures, such as quantum dots, we need to use other type of potential coupling such as {the scalar
 potential coupling \cite{ber}. 
 However, in our present work we ensure confinement of our fermions
 in the $y$-direction by using infinite mass confinement, which requires
 infinite mass at the boundary of the $y$-strip and results in a specific
 quantization of the $y$-component of the momentum \cite{ber}.}

 {For the solution of the electrostatic problem at hand,
 we proceed in two complementary ways to study the tunneling of Dirac fermions through
 a biased graphene strip. First, we implement our recent developed Poincar\'e map~\cite{bahlouli},
 which is very handy and efficient for numerical computation. Second, we use an analytical approach
 to solve the effective 1D Dirac equation in the presence of an electrostatic barrier. Comparison
 between the results generated by both approaches shows  complete agreement.} 

The paper is organized as follows. In section 2, we describe our theoretical model Hamiltonian and apply the Poincar\'e
map approach, based on
the space discretization of the effective 1D Dirac equation. In section 3, we expose the direct analytical
approach to solve the same problem. In section 4, we proceed to discuss the numerical implementation of our
approaches to a specific model potential, the linear potential which generates a static electric field,
and make a comparative study between the two approaches.

\section{Poincar\'e map}

Before we embark on the two approaches mentioned above, we would like to describe mathematically our system
of  massless Dirac fermions within a strip of graphene characterized by  a very large length scale, and a
width $W$ in the presence of the applied linear potential $V(x)$ between $x=0$ and $x=L$. So our system is
composed of three major regions: the extremes $(\sf I)$ and $(\sf III)$ contain intrinsic graphene free of
any external potentials and an intermediate region $(\sf II)$ subject to the applied linear potential $V(x)$.
Graphene band structure has two Fermi points, each with a two-fold band degeneracy, and can be described by a
tight binding Hamiltonian describing two interlacing honeycomb sublattices. At low energies this Hamiltonian
can be can be described by a continuum approximation to the original tight binding model which reduces to the
two dimensional Dirac equation with a four-component envelope wavefunction whose components are labeled by a
Fermi-point pseudospin $=\pm 1$. Specifically, the Hamiltonian for one-pseudospin component for the present
system can be written as  
\begin{eqnarray}\label{ak}
 H=v_F \vec{\sigma}\cdot\vec{p}+V(x)
\end{eqnarray}
where $v_F\simeq9.84\times10^{6}m/s$ is the Fermi velocity
and $\vec{\sigma}=(\sigma_{x},\sigma_{y})$ are the Pauli matrices. 
Hereafter we set our units such that $v_F=\hbar=1$. The linear potential $V(x)$ has the following form
\begin{equation} \label{vxx}
V(x)=\left\{\begin{array}{cc} {-Fx+V_0}, & \qquad {0<x<L} \\ {0}, & \qquad {\mbox{otherwise}} \end{array}\right.
\end{equation}
where
$F=\frac{V_0}{L}$ is the strength of the static electric field. This potential configuration is shown in Figure~1
below.
\begin{center}
\includegraphics [width=3in,keepaspectratio]
{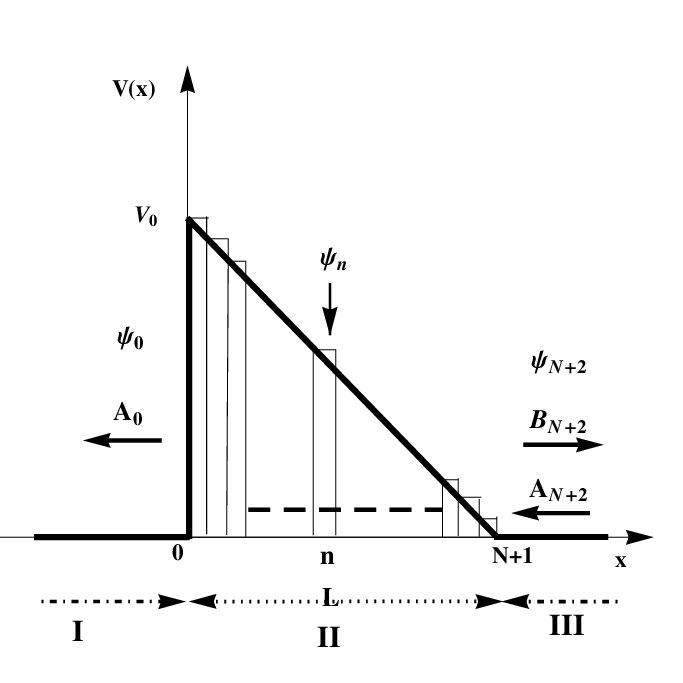}\\
{\sf{Figure 1: Discretization of the linear potential $V(x)$. }}
\end{center}

Our system is supposed to have a finite width $W$ with infinite mass boundary conditions for
the wavefunction at the boundaries $y = 0$ and $y = W$ along the $y$-direction~\cite{ben,ber}.
This boundary condition results in a quantization of the transverse momentum along the $y$-direction, which gives
\begin{equation}\lb{1}
k_{y}=\frac{\pi}{W}\left(l+\frac{1}{2}\right), \qquad l=0,1,2\cdots.
\end{equation}
One can therefore assume a spinor solution of the following form  $\psi^{j}(x,y)=\left(\phi_{{1}}^{j}(x),\phi_{{2}}^{j}(x)\right)^{\dag}e^{ik_{y}y}$ where the superscript  $j={\sf I}, {\sf II}, {\sf III}$, indicates the space region while the subscripts indicate the two spinor components. 
Thus our problem reduces to an effective 1D problem whose Dirac equation can be written as
\begin{equation} \label{1d}
\left(\begin{array}{cc} {V(x)-\varepsilon } & {\frac{d}{dx} +k_{y}} \\ {-\frac{d}{dx} +k_{y}} & {V(x)-\varepsilon } \end{array}\right)\left(\begin{array}{c} {\phi_{{1}}^{j}(x)} \\ {} \\ {-i \phi_{{2}}^{j}(x) } \end{array}\right)=0.
\end{equation}
Due to the space dependence of the potential $V(x)$ we make the following transformation on our spinor components to enable us to obtain Schrodinger like equations for each component,  $\chi_{{1}}^{j}=\frac{1}{2}\left(\psi_{{1}}^{j}+\psi_{{2}}^{j}\right)$ and $\chi_{{2}}^{j}=\frac{1}{2i}\left(\psi_{{1}}^{j}-\psi_{{2}}^{j}\right)$, which obey the  coupled stationary equations. These are
\begin{eqnarray}\label{co}
\frac {d\chi_{{1,2}}^{j}(x)}{dx}\pm i \left( V(x)-\epsilon \right)\chi_{{1,2}}^{j}(x)\mp ik_{{y}}\chi_{{{2,1}}}^{j}(x) =0.
 \end{eqnarray}
Each spinor component $\chi_{{1,2}}^{j}$  can be shown to satisfy the following uncoupled second order differential equation
\begin{equation}\label{df}
{\frac {d^{2}}{d{x}^{2}}}\chi_{{{1,2}}}^{j} \left( x \right) + \left( \pm i{\frac
{d}{dx}}V \left( x \right) + \left[ V \left( x \right) -\varepsilon \right] ^{2}
-{k_y}^{2} \right) \chi_{{{1,2}}}^{j} \left( x \right)=0.
\end{equation}

In this section we will apply the Poincar\'e map approach to solve the above effective 1D Dirac equation.
In this approach we start by subdividing the potential interval $L$ into $N +1$ regions (Figure 1).
In every $n$-th region we approximate the linear potential by a constant value $V_n = V(x_n)$  where $x_n = nh$
and $h=\frac{L}{N+1}$. Hence, the Dirac equation in each region ($n$), defined by $h(n - 1) <x< hn$, can be easily
solved for the piece-wise constant potential.
For simplicity, we chose the incident wave propagating from right to left and apply the continuity of the
spinor wavefunctions at the boundary separating adjacent regions. The general solutions of equation (\ref{df})
in the $n$-th region where $V(x)=V_n$ are given by 
\begin{equation} \label{GrindEQ5}
\psi_{{n}}= A_{n}\left(
                      \begin{array}{c}
                        1 \\
                        -z_{n}^{*} \\
                      \end{array}
                    \right) e^{-ik_{n}x}+B_n\left(
                                           \begin{array}{c}
                                             1 \\
                                             z_{n} \\
                                           \end{array}
                                         \right)e^{ik_{n}x}
\end{equation}
with $k_{n}=\sqrt{(\varepsilon-V_{n})^{2}-k_{y}^{2}}$, the complex number $z_{n}$ is defined by $z_{n}=\frac{1}{z_{n}^{*}}={\mbox{sgn}} \left(\varepsilon-V_n\right)\frac{k_{n}+ik_{y}}{\sqrt{k_{n}^{2}+k_{y}^{2}}}$.
In order to obtain the relationship between $\psi_{n+1}$ and $\psi_{n}$ we apply continuity of
$\psi$ at the boundary $x =x_n$ (Figure 2). This leads to 
\begin{equation} \label{GrindEQ11}
M_{n}(x_n)\left(
                  \begin{array}{c}
                    A_{n} \\
                    B_{n} \\
                  \end{array}
                \right)=M_{n+1}(x_n)\left(
                  \begin{array}{c}
                    A_{n+1} \\
                    B_{n+1} \\
                  \end{array}
                \right).
\end{equation}
Also $M_{n+1}(x_{n+1})$ and $M_{n+1}(x_n)$ are related by
\begin{equation}\label{GrindEQ__12_}
M_{n+1}(x_{n+1})=M_{n+1}(x_n) S_{n+1}, \qquad S_{n+1}=\left(
              \begin{array}{cc}
                e^{-ihk_{n+1}} & 0 \\
                0 &  e^{ihk_{n+1}} \\
              \end{array}
            \right).
\end{equation}
\begin{center}
\includegraphics [width=10.10cm,keepaspectratio]
{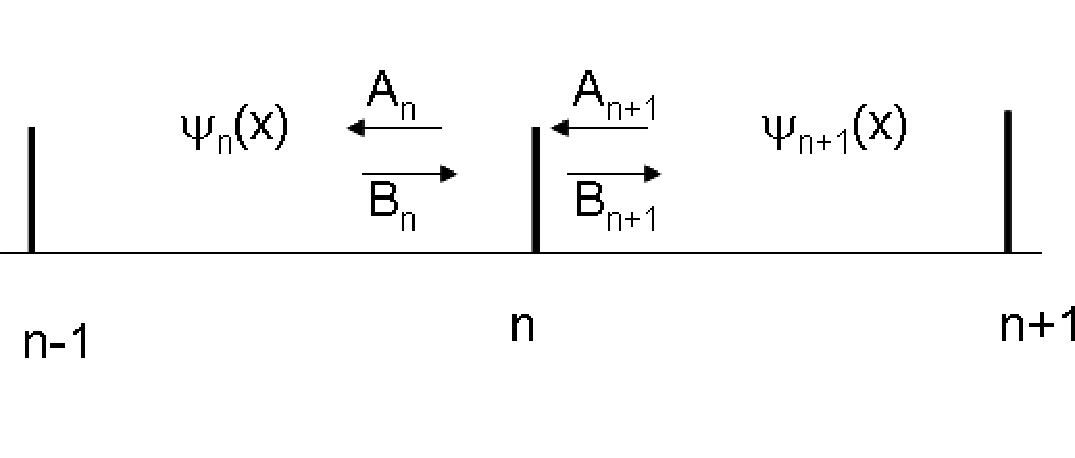}\\
{\sf{Figure 2: Solutions of the 1D Dirac equation in two consecutive regions, continuity of spinors is applied at $x=x_n$.}}
\end{center}
Using the above results we can write the desired Poincar\'{e} map as
\begin{equation}\label{GrindEQ16}
\psi_{n+1}(x_{n+1})=\tau_{n}\psi_{n}(x_n)
\end{equation}
where we have defined a simplified notation by $\psi_{n}= \psi_{n}(x_n)$ and  $\tau_{n}=M_{n+1}(x_{n+1})S_{n+1}M_{n+1}^{-1}(x_{n+1})$, or more explicitly
\begin{equation}
\tau _{n}= \frac{1} {{z_{(1+n)}^*}+z_{(1+n)}} \\
\left(
\begin{array}{cc}
 z_{(1+n)}^*e^{i k_{(1+n)}} +  z_{(1+n)} {e^{-i k_{(1+n)}} } &   -e^{-i k_{(1+n)}} +   {e^{-i k_{(1+n)}} } \\
 -e^{-i k_{(1+n)}} +   {e^{-i k_{(1+n)}} } & z_{(1+n)}^*e^{-i k_{(1+n)}} +  z_{(1+n)} {e^{-i k_{(1+n)}} }
\end{array}
\right).
\end{equation}

To make use of the above Poincar\'{e} map in solving our scattering problem we need to define our incident,
reflected and transmitted waves. For $x \leq 0$ where $V = 0$ (region {\sf I}), we can use for our transmitted spinor
evaluated at $n = 0$, the 
suitably normalized form 
\begin{equation}\label{GrindEQ6}
    \psi_{0}=\left(
               \begin{array}{c}
                 1 \\
                 -{z_{0}^*} \\
               \end{array}
             \right).
\end{equation}
This is juste the value of the transmitted wave at the zeroth site, $x = 0$ $(n = 0)$,
which is given by
\begin{equation}\label{GrindEQ7}
    \psi_{L}=A_{0} \left(
               \begin{array}{c}
                 1\\
                 -{z_{0}^*} \\
               \end{array}
             \right)e^{-ik_{0}x}.
\end{equation}
On the other side, for $x \geq h(N+1)$ where $ V = 0 $ (region {\sf III}), we have both incident and reflected spinor
waves. Just outside the potential region on the right hand side in the $(N+2)$-th region the spinor wave
can be written as
\begin{equation}\label{GrindEQ8}
  \psi_{R}=  A_{N+2}\left(
                      \begin{array}{c}
                       1 \\
                        -{z_{0}^*} \\
                      \end{array}
                    \right) e^{-ik_{0}x}+B_{N+2}\left(
                                                  \begin{array}{c}
                                                    1 \\
                                                     z_{0} \\
                                                  \end{array}
                                                \right)e^{ik_{0}x}.
\end{equation}
Hence to evaluate the transmission amplitude all we need is to find $A_{N+2}$ using the above recursive
scheme. Our strategy now is to express $A_{N+2}$ in terms of $\psi_{N+1}$ and $\psi_{N+2}$, the two end point spinors.
This can be easily done using our previous relationships and leads to
\begin{equation}\label{GrindEQ21}
A_{N+2}=\frac{e^{ihk_{0}(N+2)}}{2(1-e^{2ihk_{0}})}
\left(
 \begin{array}{ccc}
 1 & & -z_{0} \\
 \end{array}
 \right) \left(\psi_{N+2}-e^{ihk_{0}}\psi_{N+1}\right).
\end{equation}
From the above notation we can easily define the transmission amplitude as follows
\begin{equation}\label{GrindEQ9}
t=\frac{1}{A_{N+2}}.
\end{equation}
Summing up, our numerical procedure requires first that we iterate the Poincar\'{e} map  (\ref{GrindEQ16}) to obtain the end point spinors, $\psi_{N+1}$ and $\psi_{N+2}$, in terms of the normalized transmitted spinor. These spinors will then be injected in  (\ref{GrindEQ21}) and  (\ref{GrindEQ9}) to determine the transmission amplitude. The transmission coefficient is given by $T=\left|t\right|^{2}$. The numerical implementation of this scheme in the case of linear vector potential will be done in section 4.

{Before closing this section, we would like to point out that transfer matrix methods have been used heavily
in the context of transport in graphene \cite{mr} and graphene superlatices \cite{bliokh}. However, the Poincar\'e map,
which can be of great interest in application related to disordered choatic systems, applies only to discretized systems
and has been applied in its present form only recently to the Dirac equation \cite{bahlouli}.}

\section{Analytical method}

Let us now solve analytically the effective 1D Dirac equation or equivalently equation (\ref{df}) in
the presence of  {an electrostatic barrier} (region {\sf II}). Our objective is to find the transmission coefficient
for a Dirac fermion scattered  by a linear potential and then compare our results with those found in
previous section using the Poincar\'e map method.
{Before we proceed further, we would like to mention that the transmission through a trapezoidal
barrier in graphene was analytically calculated by Sonin \cite{sonin}. However, the exact solution
of the Dirac equation in uniform electric
field in terms of confluent hypergeometric functions was found long time
ago by Sauter \cite{sauter}.}

The solution of equation (\ref{df}) in region {\sf I}
and {\sf III}  are given by
\begin{equation}
\phi^{{\sf I}}(x)= \left(%
\begin{array}{c}
  1 \\
  z \\
\end{array}%
\right)e^{ik_{x}x}+r\left(%
\begin{array}{c}
  1 \\
  -z^{\ast} \\
\end{array}%
\right) e^{-ik_{x}x}, \qquad \phi^{\sf III}(x)=t\left(%
\begin{array}{c}
  1 \\
  z \\
\end{array}%
\right)e^{ik_{x}x}
\end{equation}
where $r$ and $t$ are the reflection and transmission amplitudes, respectively.
The wave vector $k_{x}=\sqrt{\varepsilon^{2}-k_y^{2}}$ and the complex number $z$ is defined by $z={\mbox{sgn}} (\varepsilon) (k_{x}+ik_{y})/\sqrt{k_{x}^{2}+k_{y}^{2}}$. In region {\sf II} the general solution can be expressed in terms of the parabolic cylinder function~\cite{Ab,vi} as
\begin{equation}\label{aaaa}
\chi_{{1}}^{{\sf II}}(x)=\alpha D_{\nu-1}\left(\sqrt{\frac{2}{F}}e^{i\pi/4}(F x+E)\right)+\beta D_{-\nu}\left(-\sqrt{\frac{2}{F}}e^{-i\pi/4}(F x+E)\right)
\end{equation}
where 
$\nu=\frac{ik_{y}^{2}}{2F}$, $E=\varepsilon-V_0$,
$\alpha$ and $\beta$ are constants. Substituting  (\ref{aaaa}) in (\ref{co}) gives the other component
\begin{eqnarray}\label{cc}
\chi_{{2}}^{{\sf II}}(x)&=- \frac{\beta}{k_y}\left[2(E+Fx) D_{-\nu}\left(-\sqrt{\frac{2}{F}}e^{-i\pi/4}(F x+E)\right)+\sqrt{2 F}e^{i\pi/4}D_{-\nu+1}\left(-\sqrt{\frac{2}{F}}e^{-i\pi/4}(F x+E)\right)\right]\nonumber\\
&- \frac{\alpha}{k_y}\sqrt{2 F}e^{-i\pi/4} D_{\nu}\left(\sqrt{\frac{2}{F}}e^{i\pi/4}(F x+E)\right).~~~~~~~~~~~~~~~~~~~~~~~~~~~~~~~~~~~~~~~~~~~~~~~~~~~~~~~~~~~~~~
\end{eqnarray}
The components of the spinor solution of the Dirac equation (\ref{ak}) in region {\sf II} can be obtained from  (\ref{aaaa}) and (\ref{cc}) where $\phi_{{1}}^{{\sf II}}(x)=\chi_{{1}}^{{\sf II}}+i\chi_{{2}}^{{\sf II}}$ and $\phi_{{2}}^{{\sf II}}(x)=\chi_{{1}}^{{\sf II}}-i\chi_{{2}}^{{\sf II}}$. This results in
\begin{equation}
\psi^{{\sf II}}(x)= \alpha\left(%
\begin{array}{c}
  a^{+}(x) \\
  a^{-}(x) \\
\end{array}%
\right)+\beta\left(%
\begin{array}{c}
  b^{+}(x) \\
  b^{-}(x) \\
\end{array}%
\right)
\end{equation}
where the function $a^{\pm}(x)$ and $b^{\pm}(x)$  are given by
\begin{eqnarray}\label{}
a^{\pm}(x)&=&D_{\nu-1}\left(\sqrt{\frac{2}{F}}e^{i\pi/4}(F x+E)\right)\mp\frac{\sqrt{2 F}}{k_y}e^{i\pi/4} D_{\nu}\left(\sqrt{\frac{2}{F}}e^{i\pi/4}(F x+E)\right)\nonumber\\
b^{\pm}(x)&=&\pm\frac{1}{k_y}\sqrt{2 F}e^{-i\pi/4}D_{-\nu+1}\left(-\sqrt{\frac{2}{F}}e^{-i\pi/4}(F x+E)\right)\nonumber\\
&&\pm\frac{1}{k_y}(-2iE\pm k_y-2iFx)D_{-\nu}\left(-\sqrt{\frac{2}{F}}e^{-i\pi/4}(F x+E)\right).
\end{eqnarray}
The coefficients $r$, $\alpha$, $\beta$ and $t$ are determined from the continuity of the spinor wavefunctions at the boundaries $x = 0, L$, that is $\psi^{\sf I}(x=0)=\psi^{\sf II}(x=0)$ and $\psi^{\sf II}(x=L)=\psi^{\sf III}(x=L)$. The transmission coefficient through the linear potential is obtained from $T=\left|t\right|^{2}$ where  the corresponding amplitude $t$ is obtained from the aforementioned boundary conditions.
 It is given by
\begin{eqnarray}\label{tt}
t=\frac{e^{-ik_{x}L}\left[1+z^{2}\right]\left[b^{+}(L)a^{-}(L)-b^{-}(L)a^{+}(L)\right]}
{\left[b^{+}(0)+z b^{-}(0)\right]\left[a^{-}(L)-z a^{+}(L)\right]-
\left[a^{+}(0)+z a^{-}(0)\right]\left[b^{-}(L)-z b^{+}(L)\right]}.
\end{eqnarray}

\section{Results and discussion}

In this section we implement our previous Poincar\'e map and analytical approaches to a nanoribbon system subject to  an electric potential of strength $V_0 = 10, 20$ and a field region of length $L = 3, 10$ so that the resulting static electric field strength is given by $F = V_0/L = 10/3, 2$, respectively. In Figure 3 we show the transmission as a function of energy for a transverse momentum $k_y = 1$. The solid lines corresponds to the exact transmission derived in section 3 and given by equation (\ref{tt}) while the dashed lines are generated by our Poincar\'e map for $N = 200$ iterations, the agreement is just perfect.\\
\begin{center}
\includegraphics [width=6.6cm,keepaspectratio]
{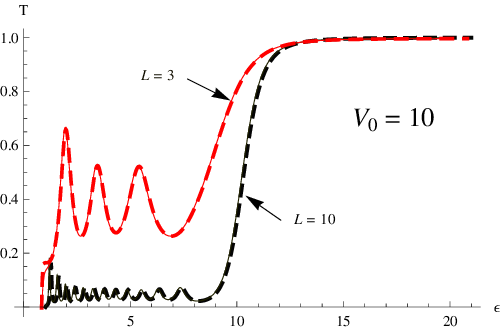}~~~~~~~~~~~~
\includegraphics [width=6.6cm,keepaspectratio]
{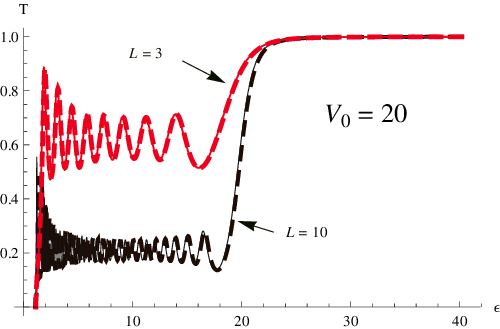}
\end{center}
\begin{center}
{\sf{Figure 3:Transmission coefficients $T$ versus energy $ \varepsilon $for
 $L = 3,10 $,  $V_0 = 10,20 $ and $k_y = 1$.
 }}
\end{center}

 \noindent {{Figure 3}} shows the concordance between the results generated by the analytical and Poincar\'e map method we adapted. We note that below a certain critical energy  $\varepsilon = k_y$ the transmission is almost zero, then it starts oscillations whose frequency increases with $L$, the size of the region subject to the electric field.
The transmission increases with $L$ and reaches unity for energies above $V_0 + 2 k_y$.

\begin{center}
\includegraphics [width=6.6cm,keepaspectratio]
{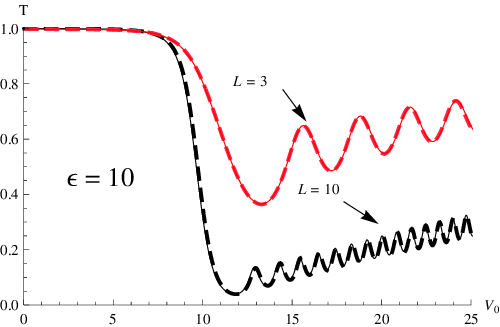}~~~~~~~~~~~~
\includegraphics [width=6.6cm,keepaspectratio]
{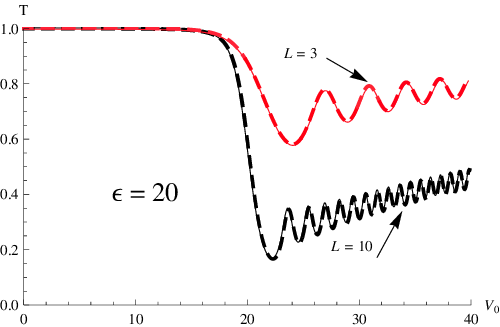}
\end{center}
\begin{center}
{\sf{Figure 4:Transmission coefficients $T$ versus $ V_0 $ for
 $L = 3,10 $, $\varepsilon = 10,20 $ and $k_y = 1 $}}
\end{center}
\noindent {{Figure 4}} shows the transmission as a function of the strength of the applied voltage, total transmission is observed for small values of $V_0$ less than the energy of the incident fermion. It then decreases sharply for $V_0 > \varepsilon - 2 k_y $ until it reaches a relative minimum and then begins to increase in an oscillatory manner.
We notice in both Figures 3 and 4 that the amplitude of oscillations and period increase as we decrease the size of the electric field region, $L$.
\begin{center}
\includegraphics [width=6.6cm,keepaspectratio]
{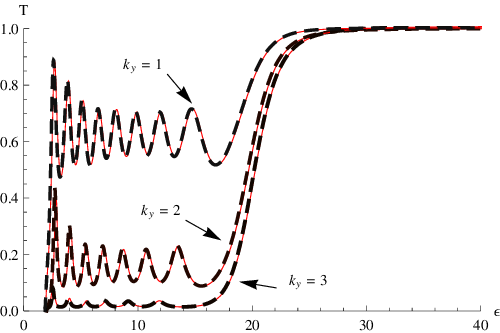}
\end{center}
\begin{center}
{\sf{Figure 5:Transmission coefficients $T$ versus energy $ \varepsilon $ for
 $L = 3 $, $V_0 = 20 $ and different values of $k_{y}$.}}
\end{center}
{Figure 5 shows that the effect of the transverse momentum $k_y$ on transmission, is antagonistic to that of length $L$. But it should be pointed out that the number of oscillations increases as $k_y$ decreases and the curves for different values of $k_y$ do not intersect.}

Now we would point out that our effective 1D massless Dirac equation is equivalent to a massive one with an effective mass equal to the transverse quantized wave vector $k_{y}$. For this purpose we would like to consider a unitary transformation, which enable us to map the effective 1D (equation (\ref{1d})) into a 1D massive Dirac equation. Such a unitary transformation does not affect the energy spectrum or the physics of the problem. We choose a rotation by
{$\frac{\pi}{2}$} about the $y$-axis, $U=e^{i{\textstyle\frac{\pi }{4}} \sigma _{2} } $. Thus, the transformed Hamiltonian and wavefunction
read 
\begin{eqnarray} \label{masse}
\left(\begin{array}{cc} {V(x)-\varepsilon+k_y } & {\frac{d}{dx}} \\ {-\frac{d}{dx}} & {V(x)-\varepsilon-k_y } \end{array}\right)\left(\begin{array}{c} {\tilde{\psi}_{{1}}^{j}(x)} \\ {} \\ {\tilde{\psi}_{{2}}^{j}(x) } \end{array}\right)=0,  \qquad \tilde{\psi}_{j_{1,2}}(x)=U\psi_{j_{1,2}}(x)
\end{eqnarray}
which is identical to a 1D massive Dirac equation with an effective mass $m = k_y$. To check the validity of this assertion numerically we show in Figure 6 the transmission as a function of energy as generated by the exact analytical result (\ref{tt}), the Poincar\'e map (\ref{GrindEQ9}) and the 1D massive Dirac equation with an effective mass $m^{\ast}=k_y$ in (\ref{masse}). We see from this figure that the three curves coincide to the point that we cannot even  distinguish between them. This lead us to include an inset in Figure 6 showing each figure translated for ease of comparison purposes.


\begin{center}
\includegraphics [width=6.6cm,keepaspectratio]
{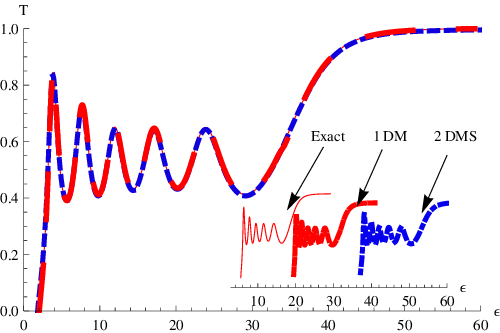}
\end{center}
\begin{center}
{\sf{Figure 6 :Transmission coefficients $T (\varepsilon)$ for
 $L = 1$, $V_0 = 40$ and $k_y = m^{\ast} = \frac{9\pi}{10}$.}}
\end{center}
This last figure confirms, numerically, the equivalence between a one-dimensional system of Dirac fermions with mass and a two-dimensional system of massless Dirac fermions constrained along the $y$-direction by an infinite mass boundary condition, hence forming a graphene nanoribbon. The transverse component of the wave vector, $k_y$, played the role of an effective mass \cite{Hai} in the resulting effective 1D  Dirac equation.

{To close this section
we would like to mention that our present work could be extended to handle a system of
2D Dirac fermions with mass $m$ as done in reference \cite{peeters}. Once confined to a strip
along the $y$-direction we will end up with an effective 1D Dirac equation with effective mass
 $m_{\sf eff}= \sqrt{m^2 +k_y^2}$. This might be considered
 as a simple extension, which is useful to model an underlying substrate.}

\section*{{Acknowledgments}}

\noindent The generous support provided by the Saudi Center for Theoretical Physics (SCTP) is highly appreciated by all Authors. AJ and (EBC, AE) acknowledge partial support by King Faisal University and KACST, respectively. We also acknowledge the support of KFUPM under project RG1108-1-2.
{We would to express our deep appreciation for the very constructive comments made by the referee}.


\begin{thebibliography}{1}

\bibitem{ag} A. K. Geim and K. S. Novoselov, Nat. Mat. 6, 183 (2007).

\bibitem{ah} A. H. C. Neto, F. Guinea,  N. M. R. Peres, K. S. Novoselov
and  A. K. Geim, Rev. Mod. Phys. 81, 109 (2009).

\bibitem{n} N. Stander, B. Huard and  D. G. Gordon, Phys. Rev. Lett. 102, 026807 (2009).

\bibitem{ka}  M. I. Katsnelson,  K. S. Novoselov and  A. K. Geim,  Nature Phys. 2 620 (2006).

\bibitem{hs} H.  Sevin\c{c}li, M. Topsakal and S. Ciraci, Phys. Rev. B 78, 245402 (2008).



\bibitem{ma} L. Dell'Anna and A. De Martino, Phys. Rev. B 79, 045420 (2009).

\bibitem{s} S. Mukhopadhyay, R. Biswas and C. Sinha, Phys. Status Solidi B 247, 342 (2010).



\bibitem{ben} J. Tworzydlo, B. Trauzettel, M. Titov, A. Rycerz and C. W. J. Beenakker,  Phys. Rev. Lett. 96, 246802 (2006).

\bibitem{ad} A. D. Alhaidari, H. Bahlouli and A. Jellal, Relativistic Double Barrier Problem with Three
Sub-Barrier Transmission Resonance Regions, {\sf arXiv:1004.3892}.

\bibitem{nov} K. S. Novoselov,  E. McCann,  S. V. Morozov,  V. I. Falko,  M. I. Katsnelson, U. Zeitler, D. Jiang,
F. Schedin and A. K. Geim,  Nature Phys. 2, 177 (2006).

\bibitem{mr}  M. R. Masir, P. Vasilopoulos  and   F. M. Peeters, New J. Phys. 11, 095009 (2009).

\bibitem{ch} E. B. Choubabi, M. El Bouziani and A. Jellal, Int. J. Geom. Meth. Mod. Phys.7, 909 (2010).

\bibitem{mo} A. Jellal and A. El Mouhafid, J. Phys. A: Math. Theo. 44, 015302 (2011).





\bibitem{ber}  M. V. Berry and R. J. Modragon,  Proc. R. Soc. London Ser. A 412, 53 (1987).

\bibitem{bahlouli} H. Bahlouli, E. B.  Choubabi and A. Jellal, Solution of One-dimensional Dirac Equation via Poincar\'e Map,
{\sf arXiv:1105.4741}, to appear in Europhys. Lett (2011).

{\bibitem{bliokh} Y.P. Bliokh, V. Freilikher, S. Savel\'ev and F. Nori, Phys. Rev. B 79, 075123 (2009).

\bibitem{sonin} E.B. Sonin, Phys. Rev. B 79, 195438 (2009).

\bibitem{sauter} F. Sauter, Zeitschrift f\"ur Physik 69, 742 (1931).}

\bibitem{Ab} M. Abramowitz and  I. Stegum, Handbook of Integrabls, Series and Products, (Dover, New York, 1956).

\bibitem{vi} L. Gonzalez-Diaz and  V. M. Villalba, Phys. Lett. A 352, 202 (2006).

\bibitem{Hai} A. D. Alhaidari, A. Jellal, E. B. Choubabi  and H. Bahlouli, Mass Generation via Space Compactification
in Graphene, {\sf arXiv:1010.3437}.

{\bibitem{peeters} M. Barbier, F.M. Peeters, P. Vasilopoulos and J. Milton Pereira, Phys. Rev. B 77, 115446 (2008).}


\end{thebibliography}
\end{document}